\documentclass[epj]{svjour}
\usepackage[utf8]{inputenc}
\usepackage{epsfig}
\usepackage[T1]{fontenc}
\usepackage{graphicx}
\usepackage{amssymb}
\usepackage{amsmath}

\usepackage{relsize}
\usepackage{color}
\usepackage{bm}
\usepackage{array}
\newcommand{\relphantom}[1]{\mathrel{\phantom{#1}}}

\begin{document}

\title{The phase diagram of random Boolean networks with nested canalizing
  functions}

\author{Tiago P. Peixoto}
\institute{Institut für Festkörperphysik, TU Darmstadt, Hochschulstrasse 6,
  64289 Darmstadt, Germany \\ \email{tiago@fkp.tu-darmstadt.de}}
\date{\today}


\abstract{
  We obtain the phase diagram of random Boolean networks with nested canalizing
  functions. Using the annealed approximation, we obtain the evolution of the
  number $b_t$ of nodes with value one, and the network sensitivity $\lambda$,
  and we compare with numerical simulations of quenched networks. We find that,
  contrary to what was reported by Kauffman \textit{et al.} [Proc. Natl.
  Acad. Sci. 2004 {\bf 101} 49 17102-7], these networks have a rich phase
  diagram, were both the "chaotic" and frozen phases are present, as well as an
  oscillatory regime of the value of $b_t$. We argue that the presence of only
  the frozen phase in the work of Kauffman \textit{et al.} was due simply to the
  specific parametrization used, and is not an inherent feature of this class of
  functions. However, these networks are significantly more stable than the
  variants where all possible Boolean functions are allowed.
}

\maketitle

\section{Introduction}

Boolean networks (BN) were introduced by
Kauffman~\cite{kauffman_homeostasis_1969,kauffman_metabolic_1969} as a simple
model of gene regulation. In this model the transcription states of the genes
are described by Boolean variables, and their dependency to other genes by
Boolean functions. In Kauffman's original model, which is usually called a
Random Boolean Network (RBN)~\cite{drossel_random_2008}, both the functions and
their inputs are randomly distributed among all possible choices. Since this
clearly discards any possible structure which may be selected by the
evolutionary process, these networks are \emph{null models} of gene regulation,
from which general features may be derived, which are independent of the missing
details~\cite{bornholdt_systems_2005}. Indeed this simple model already shows an
emergent behaviour which may be applicable to real system, namely the existence
of two dynamical phases: Frozen and ``chaotic''. In the frozen phase, small
perturbations have a limited propagation, and eventually stop. In the chaotic
phase, small perturbations propagate exponentially fast, often reaching a finite
portion of the system. It has been argued~\cite{kauffman_homeostasis_1969} that
real systems may share features with RBNs which lie exactly at the interface
between these two phases --- the so-called critical networks. In this point of
the configuration space, small perturbations propagate only linearly; thus the
system retains some stability of the frozen phase, as well as some of the
excitability of the chaotic phase, which may be necessary for the system to
respond to external signals. Although this a very interesting feature, a more
plausible comparison with real gene regulation can only be made if more
realistic properties are included, such as more realistic
topologies~\cite{fox_topology_2001,aldana_natural_2003,aldana_boolean_2003,castro_e_silva_scale-free_2004,serra_dynamics_2004,kinoshita_robustness_2008,drossel_critical_2009}
or update
functions~\cite{rohlf_criticality_2002,moreira_canalizing_2005,szejka_phase_2008},
for instance. In this paper we will consider RBNs with nested canalizing
functions (NCFs), introduced in
\cite{kauffman_random_2003,kauffman_genetic_2004}. These functions are a natural
extension of the concept of \emph{canalization} often present in biological
systems~\cite{waddington_canalization_1942,harris_model_2002}. A function with a
canalizing input is such that if this input is at its canalizing value (either
$1$ or $0$), then the output of the function is automatically defined, for any
combination of the remaining input values. It has been observed that many real
functions have a canalizing input~\cite{harris_model_2002}. If this concept is
carried out to the remaining inputs, such that a hierarchy of canalization is
present, the resulting function is a nested canalizing function. Interestingly,
the majority of the real functions studied in~\cite{harris_model_2002} are also
NCFs~\cite{kauffman_random_2003}. In this work we will obtain the phase diagram
of RBNs with NCFs. Contrary to what was claimed in~\cite{kauffman_genetic_2004},
such networks possess a rich phase diagram, where both the chaotic and frozen
phases occupy sizable portions of the configuration space. We also observe
oscillations in the number of values of $1$'s in the network, for a portion of
the parameter space.

This paper is divided as follows. In Sec.~\ref{sec:model} we define the model,
as well as nested canalizing functions, and review some known facts. In
Sec.~\ref{sec:bt} we obtain the evolution of fraction $b_t$ of $1$'s, and in
Sec.~\ref{sec:lambda} we obtain the network sensitivity $\lambda$ and the phase
diagrams. We then conclude in Sec.~\ref{sec:conclusion}, and provide some final
considerations.

\section{The Model}\label{sec:model}

A BN is defined as a directed network of $N$ nodes representing
Boolean variables $\mathbf{\sigma} \in \{1,0\}^N$, which are subject to a
dynamical update rule,
\begin{equation}
  \sigma_i(t+1) = f_i\left(\mathbf{\sigma}(t)\right)
\end{equation}
where $f_i$ is the update function assigned to node $i$, which depends
exclusively on the states of its inputs. In this work we consider that all nodes
are updated in parallel. All nodes have the same number of inputs $k$ which are
chosen randomly.

Starting from a random configuration, the dynamics of the system evolves and
eventually settles on an attractor, after a transient time. We will characterize
the properties of the system after this transient time by the fraction $b_t$ of
the number of $1$'s in the network, and by the network sensitivity $\lambda$,
which will differentiate between the dynamical phases.

\subsection{Nested canalizing functions}\label{sec:ncf}

As introduced in~\cite{kauffman_random_2003}, a nested canalizing function
$f(\{\sigma_i\})$, with inputs $\{\sigma_i\}$, $i\in[0..k-1]$ , is defined as
\begin{equation}\label{eq:ncf}
  f(\{\sigma_i\}) =
  \begin{cases}
    s_0 & \text{if } \sigma_0 = c_0 \\
    s_1 & \text{if } \sigma_0 \ne c_0 \text{ and } \sigma_1 = c_1 \\
    s_2 & \text{if } \sigma_0 \ne c_0 \text{ and } \sigma_1 \ne c_1 \text{ and } \sigma_2 = c_2 \\
    \vdots \\
    s_{k-1} &  \text{if } \sigma_0 \ne c_0 \text{ and } \dots \text{ and } \sigma_{k-1} = c_{k-1} \\
    s_d  & \text{otherwise},
  \end{cases}
\end{equation}
where $c_i\in[0,1]$ is the canalizing value of input $i$ and, and $s_i\in[0,1]$
is the output value associated with input $i$. The value $s_d$ is the default
output of the function, when no input is at its canalizing value. It is usually
assumed that $s_d = 1 - s_{k-1}$, so that every input is sensitive. However, in
this paper we will also consider the situation where $s_d$ is a free
parameter. We will call an input with $s_i=1$ an \emph{activator}, and a
\emph{deactivator} otherwise; and if an input is at its canalizing value, we
will say it is canalized.

We will consider RBNs where the functions are chosen randomly from all possible
Nested Canalizing functions, which are weighted according to the following
parameters: $a$, the probability that an input is an activator (i.e. $s_i=1$),
$c$, the probability that the canalizing value of an input is $1$, and $d$, the
probability that the default output is $1$. In the situation where $s_d = 1 -
s_{k-1}$ this last parameter is omitted.

In~\cite{harris_model_2002} it was shown that many eukaryotic genes seem to be
regulated by canalizing functions, and in~\cite{kauffman_random_2003} it was
further identified that all but $6$ of the $139$ genes studied
in~\cite{harris_model_2002} are in fact regulated by NCFs.  There is a simple
interpretation for the existence of nested canalizing functions in real gene
regulation networks: Genes are used to encode mRNA via a protein called RNA
Polymerase (RNAP). This protein binds to a region of the DNA called the promoter
region, which starts shortly before the gene itself. Genes which serve as inputs
for other genes encode proteins which are called transcription factors
(TF). These proteins bind to the upstream region of the gene, i.e. the region
preceding (and including) the promoter region. The presence of a TF close to the
promoter region may increase or decrease the probability that the RNAP will
bind, and initiate the transcription. It is easy to imagine a situation where an
hierarchy of TFs exists, where a TF which binds closer to the promoter region
has more relevance, and increases or decreases the binding probability of RNAP
by such a factor that it overrides the TFs which are bound further away, giving
rise to a (nested) canalizing input.

Let us appreciate how much of a deviation is the occurrence of NCFs, in
comparison to canalizing functions on only one input, as well as to all possible
functions, by considering the total number of functions belonging to each class.
Nested canalizing functions are identical to the \emph{unate cascade functions}
known in computer science~\cite{jarrah_nested_2007}, which have optimal
properties regarding their computation time via binary decision
diagrams~\cite{butler_average_2005}, and for which many properties are
known. According to~\cite{bender_asymptotic_1978}, the number of different NCFs
with $k$ inputs scales as $N_{nc} \sim \alpha k!  2^{k(1-\log_2\ln 2)}$, for
$k\gg 1$, where $\alpha$ is some constant. For comparison, consider the number
of functions which are canalizing on at least one input~\cite{just_number_2004},
$N_c \sim 4k2^{2^{k-1}}$. Although the fraction of both these classes relative
to the total number $2^{2^k}$ of functions with $k$ inputs vanishes for large
$k$, using the Stirling approximation for $k!$, we can easily see that the
fraction of NCFs is smaller by a factor of $N_{nc}/N_c\sim1/2^{2^{k-1}}$. Thus
the presence of nested canalization is a much stronger deviation from a random
distribution than single-input canalization.

\section{The evolution of number of 1's, $b_t$.} \label{sec:bt}

In order to obtain the fraction $b_t$ of $1$'s in the network at time $t$, we
will employ the annealed approximation~\cite{derrida_random_1986}. This is a
mean-field approximation, which assumes that the inputs of each function are
randomly chosen at each time step. By construction, this forbids local
correlations from arising, which makes the analysis easier. Since a quenched
disorder should be indistinguishable from an annealed one, in the limit of large
networks, this approximation is expected to be exact for large RBNs.

\begin{figure}[hbt!]
  \includegraphics*[width=\columnwidth]{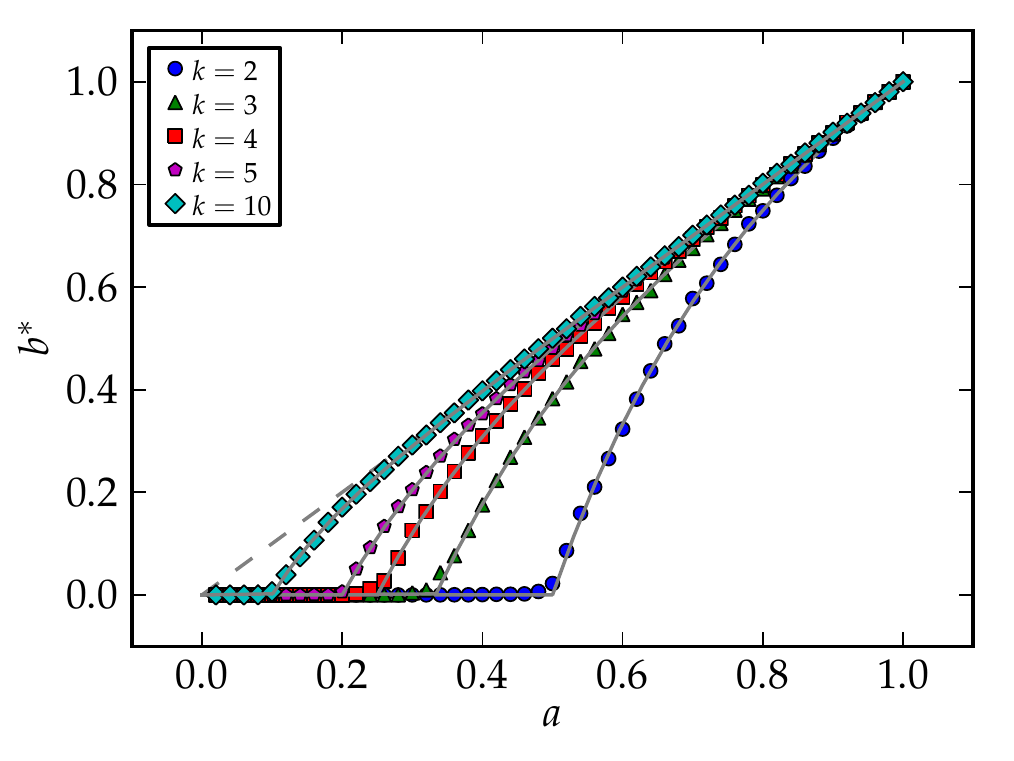}
  \includegraphics*[width=\columnwidth]{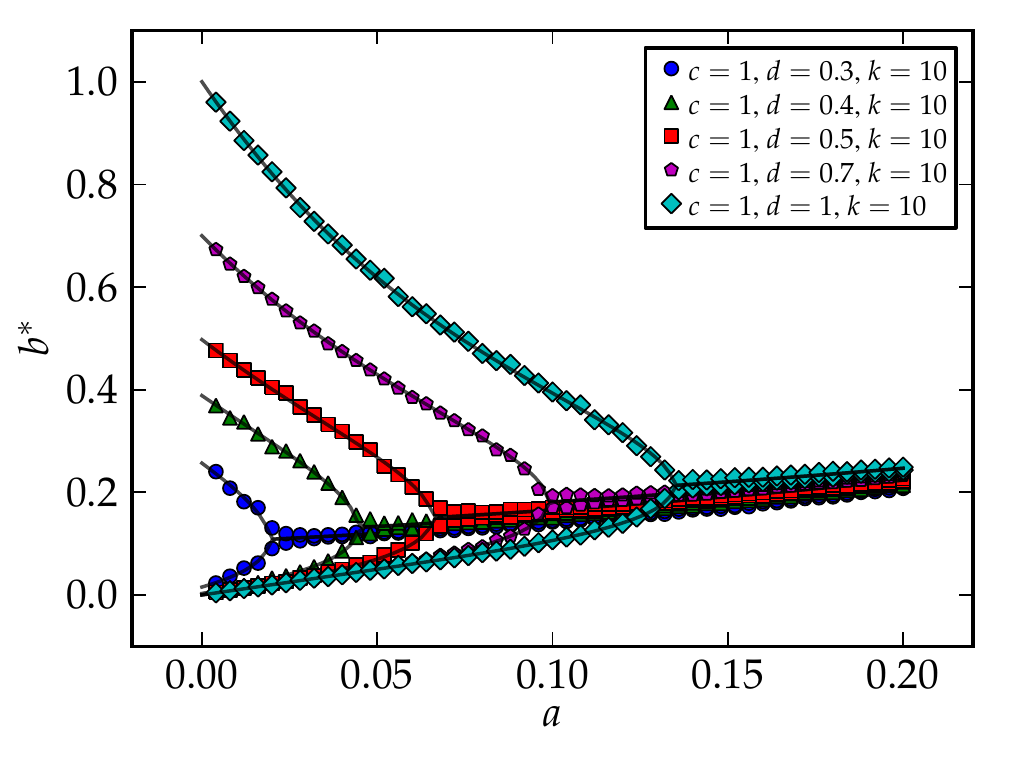}
  \caption{\label{fig:b}Fraction $b^*$ of the number of $1$'s in the network
    after the transient time, as a function of the fraction $a$ of
    activators. On top, for $c=1$ and $d=1$, we have $b^*>0$ only for $a > a^*$,
    where $a^* = 1/k$. On the bottom it is shown the bifurcation diagrams, from
    an oscillatory regime to fixed point, for the parameters indicated in the
    legend. For both figures, the symbols represent averages over $10$
    independent numerical realizations of the dynamics, for networks with
    $N=10^5$, and the solid lines are solutions of Eq.~\ref{eq:b}. }
\end{figure}

We begin by defining the probability $\gamma(b_t)$ that a random input in the
network is at its canalizing value,
\begin{equation}
  \gamma(b_t) = b_tc + (1-b_t)(1-c).
\end{equation}
Thus, the probability that the output of a randomly chosen function will be $1$
at the next time step is given simply by the probability that at least one input
is canalized and is an activator, or that the default output value is $1$,
\begin{equation}\label{eq:b}
  \begin{split}
    b_{t+1} &= (1 - (1-\gamma(b_t))^k)a + (1-\gamma(b_t))^kd \\
           &= a + (d-a)(1-\gamma(b_t))^k.
  \end{split}
\end{equation}
The situation where $s_d = 1 - s_{k-1}$ can be obtained simply by setting
$d=1-a$. The equilibrium value $b^* = b_\infty$ can then be obtained by solving
the above equation for $b_{t+1} = b_t = b^*$. We note that $\lim_{k\to\infty}
b^*=a$, except for $(a,c) \in \{(0,1), (1,0)\}$, in which case $b^*=d$.  In
Fig.~\ref{fig:b} it is shown some of the solutions as a function of $a$,
compared with numerical simulations. There are two interesting behaviors: 1. If
$c=1$ and $d=0$ (or symmetrically if $c=0$ and $d=1$, not shown), there is a
second-order transition of the value of $b^*$ at $a=1/k$ (or $a=1-1/k$), below
which $b^* = 0$ and above which $b^* > 0$. For other values of $c$ and $d$, this
behaviour is replaced by a continuous variation of $b^*$ (not shown); 2. If $|d
- c|$ is small enough, the value of $b_t$ shows an oscillatory behaviour after a
the transient time, between two values of $b^*$.  These oscillations happen when
most of the default output values correspond to the canalizing values of a large
portion of the inputs, and most canalized outputs are different from the
canalizing values. In such situation, the canalized inputs tend to deactivate
themselves, which in turn will increase the default activations, and so on. The
transition from this period-2 oscillation to a fixed point is through a
pitchfork bifurcation, where the fixed point becomes unstable ($|d b_{t+1} /
db_t(b^*)| > 1$) and gives rise to the oscillation (as shown in the bottom of
Fig.\ref{fig:b}).

\section{The network sensitivity, $\lambda$.} \label{sec:lambda}
\begin{figure*}[htb!]
  \centering
  \includegraphics*[width=0.3\textwidth]{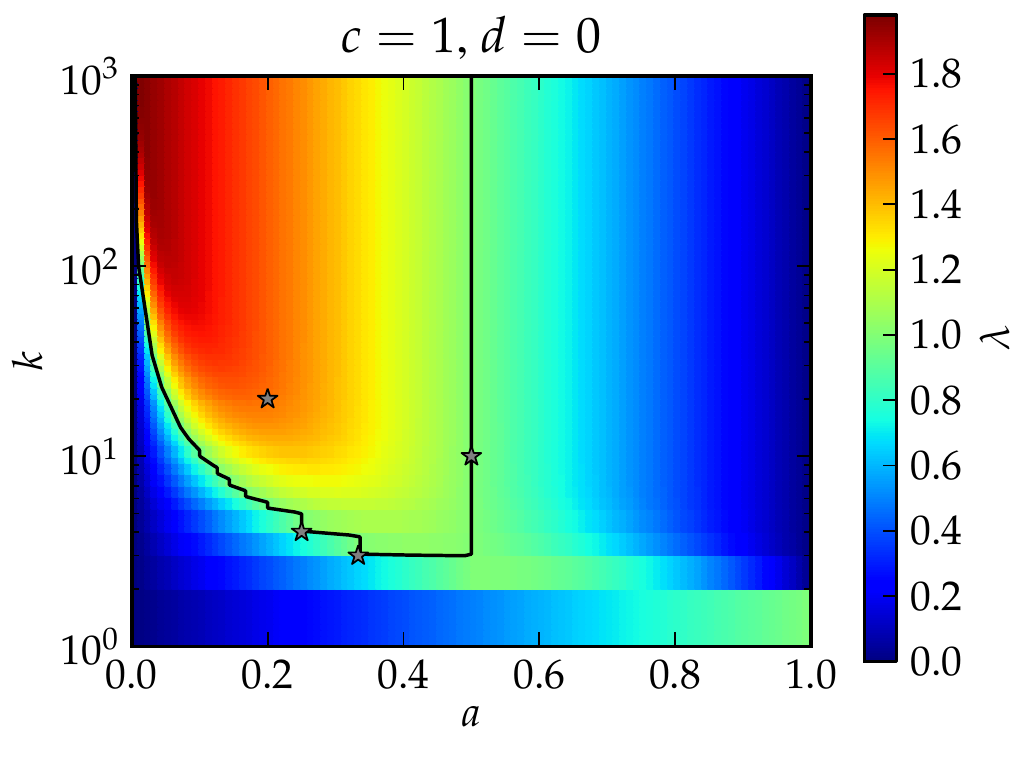}
  \includegraphics*[width=0.3\textwidth]{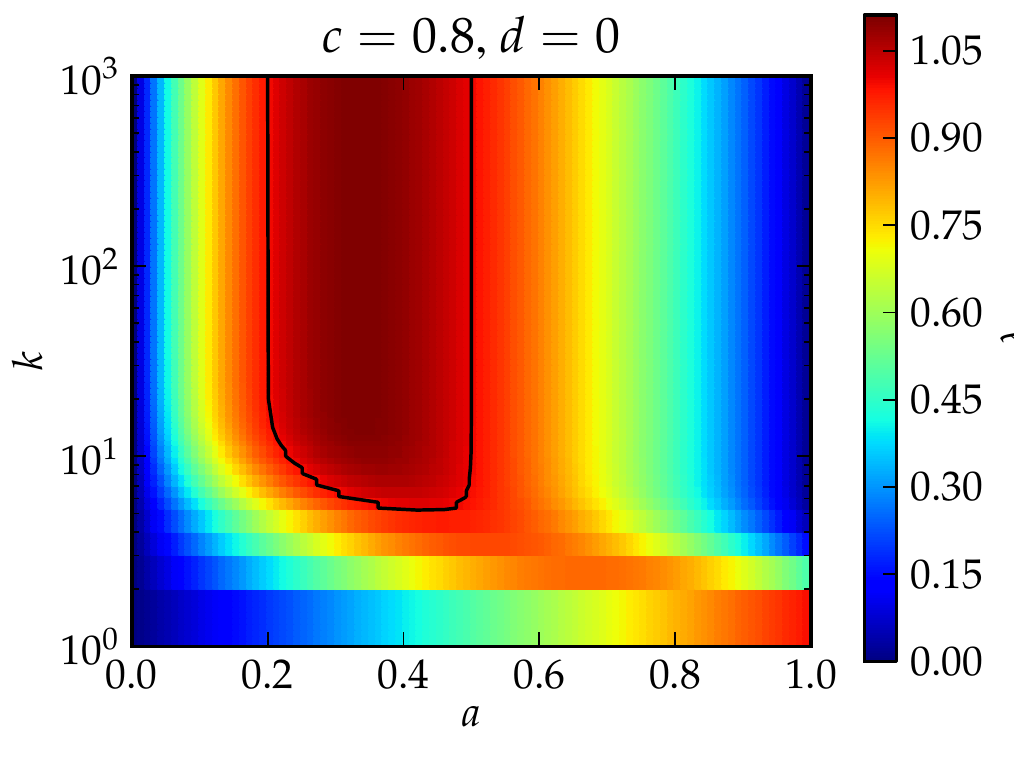}
  \includegraphics*[width=0.3\textwidth]{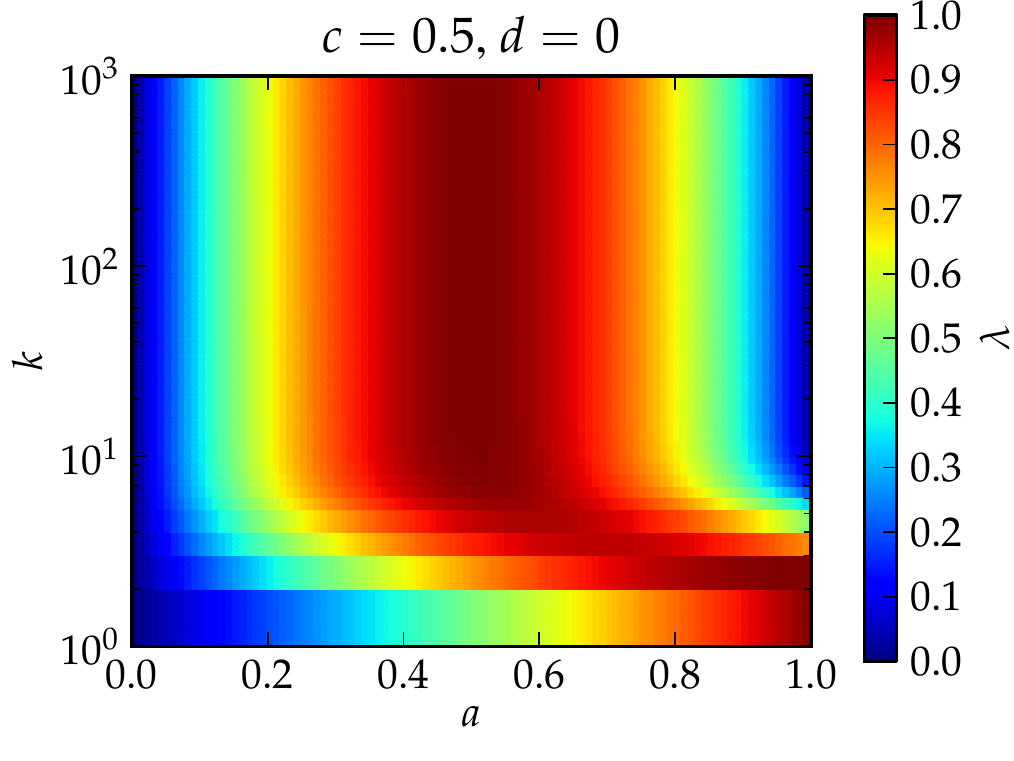}\\
  \includegraphics*[width=0.3\textwidth]{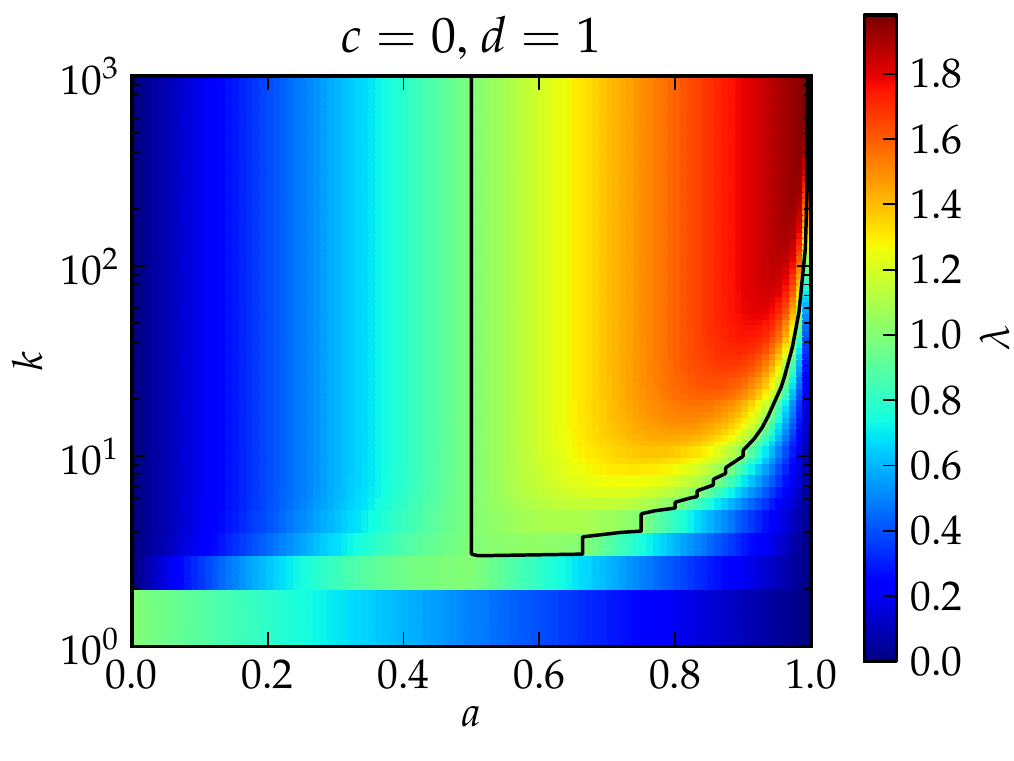}
  \includegraphics*[width=0.3\textwidth]{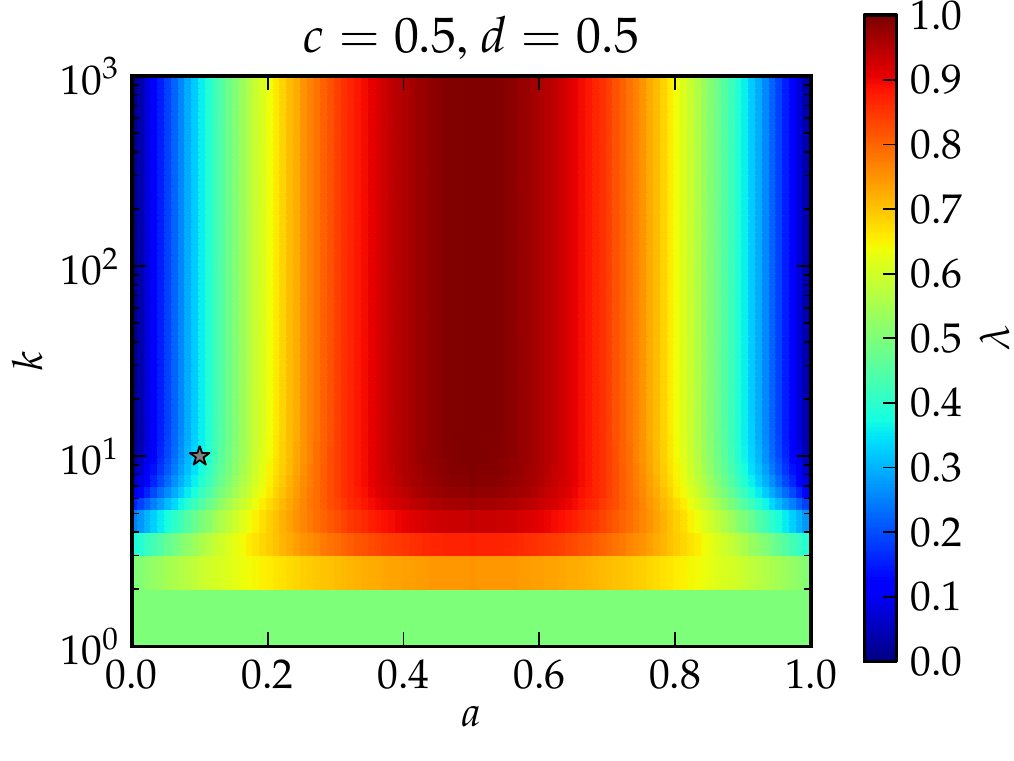}
  \includegraphics*[width=0.3\textwidth]{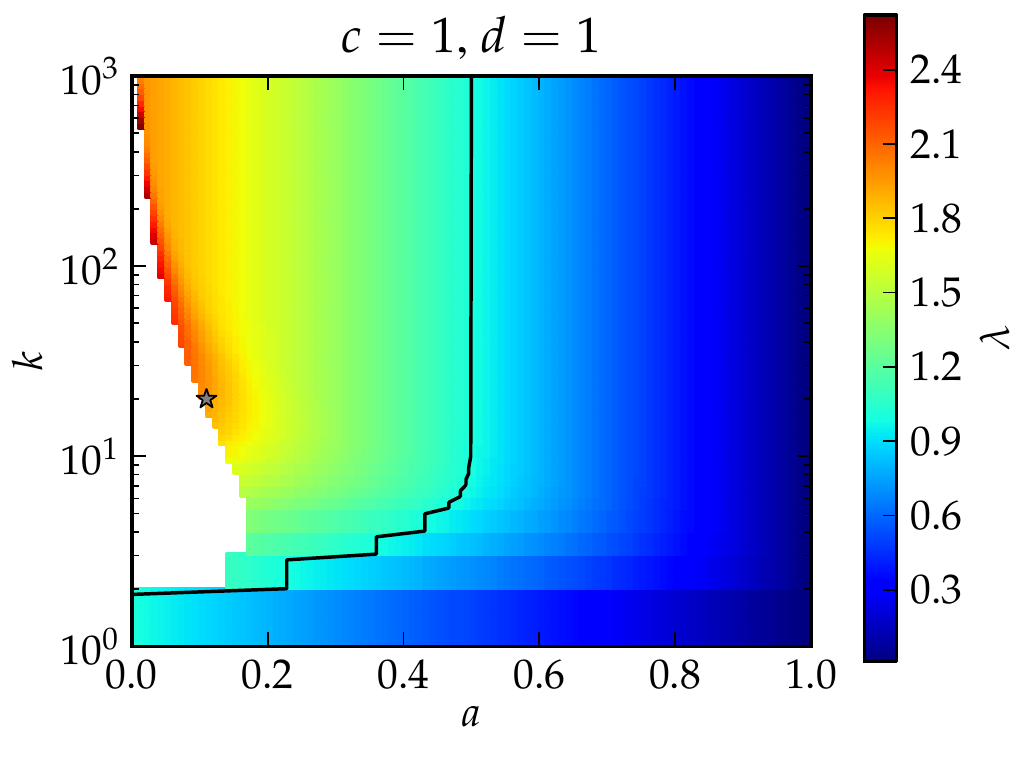}
  \caption{\label{fig:lambda} Network sensitivity $\lambda$
    (Eq.~\ref{eq:lambda}) for several parameter values, as indicated in the
    labels and axes. The critical line $\lambda = 1$ is indicated by the solid
    lines. The regions in white correspond to parameter regions where the system
    displays oscillations, and thus the value of $\lambda$ is not well
    defined. The points marked with stars ($\star$) were verified empirically in
    Fig.~\ref{fig:derrida}.}
\end{figure*}
\begin{figure*}[htb!]
  \centering
  \includegraphics*[width=0.3\textwidth]{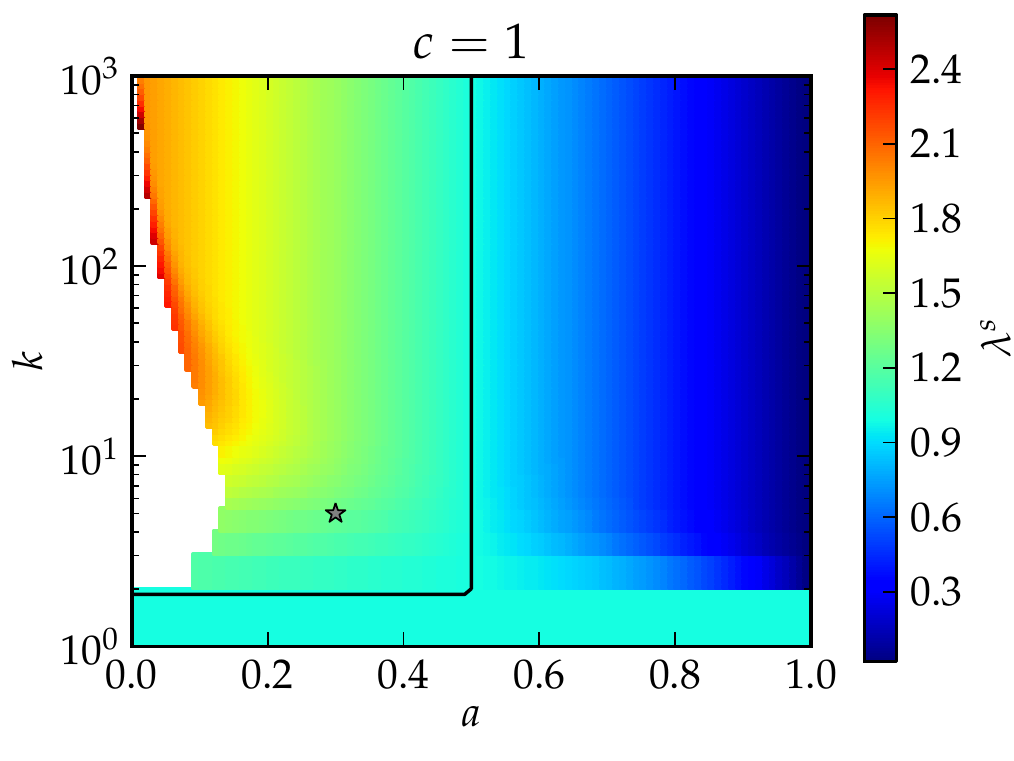}
  \includegraphics*[width=0.3\textwidth]{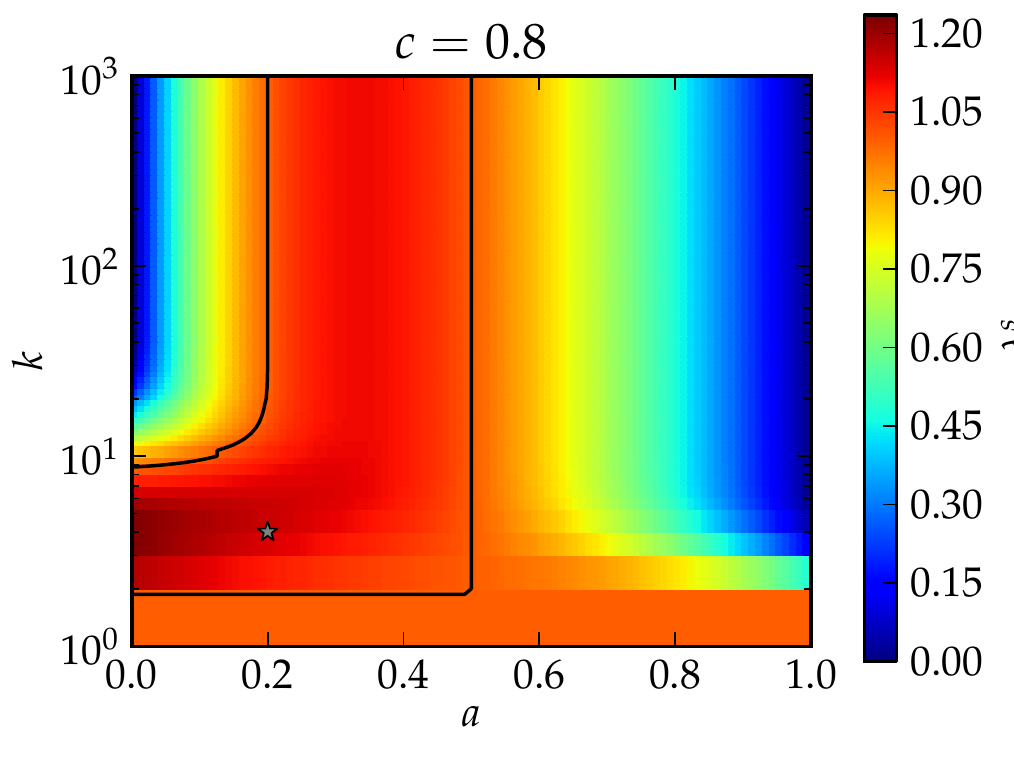}
  \includegraphics*[width=0.3\textwidth]{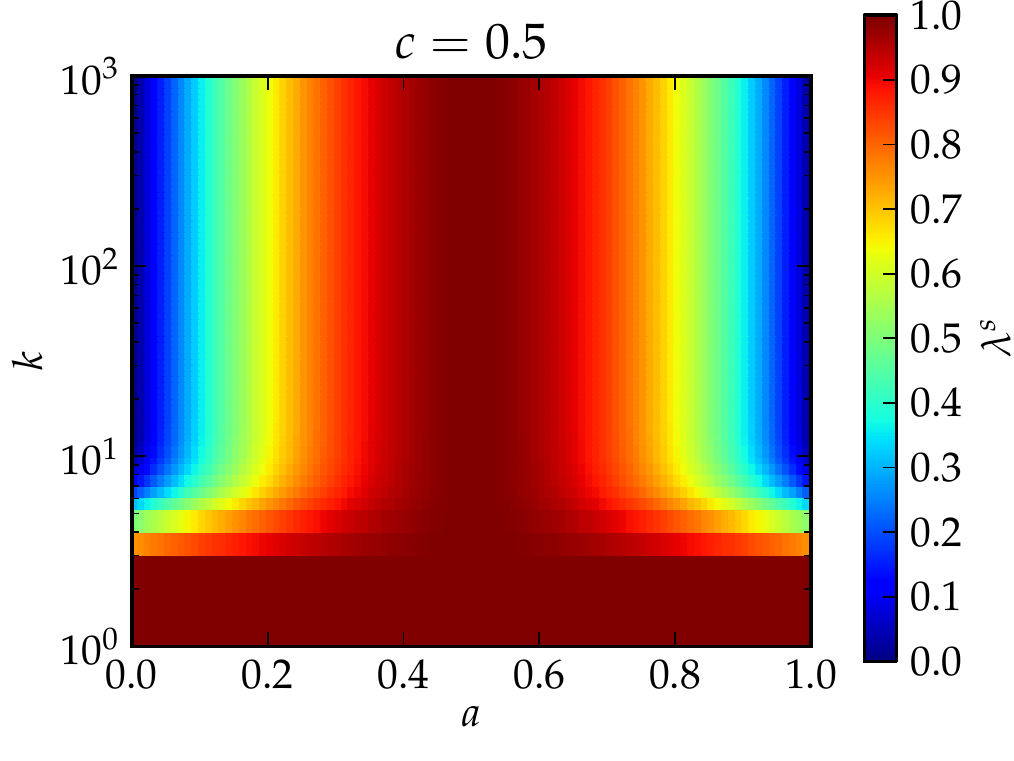}
  \caption{\label{fig:lambdas} Network sensitivity $\lambda^s$
    (Eq.~\ref{eq:lambdas}) for the case were $s_d = 1 - s_{k-1}$, for several
    parameter values, as indicated in the labels and axes. The critical line
    $\lambda^s = 1$ is indicated by the solid lines. The regions in white
    correspond to parameter regions where the system displays oscillations, and
    thus the value of $\lambda^s$ is not well defined. The points marked with
    stars ($\star$) were verified empirically in Fig.~\ref{fig:derrida}.}
\end{figure*}

The response of the system to small perturbations is characterized by the
network sensitivity
$\lambda$~\cite{luque_lyapunov_2000,shmulevich_activities_2004}, which is
defined as $k$ times the probability that the output of randomly selected
function will change if a randomly selected input is flipped. Thus, for large
networks, the average number of nodes flipped after some small time $t \ll \ln
N$ should be $\lambda^t$. Therefore if $\lambda < 1$ the number of affected
nodes will tend to zero after some time, and the dynamics is said to be in the
frozen phase. If $\lambda > 1$, the number of affected nodes will increase
exponentially, and the dynamics is said to be in the ``chaotic'' phase. For the
special value of $\lambda=1$, the average number of affected nodes increases
linearly with time, and the dynamics is said to be in the critical line between
the two phases.

We can obtain the value of $\lambda$ for RBNs with NCFs with the annealed
approximation, as we did for the values of $b^*$. We start by defining the
probability $\eta$ that two consecutive inputs have the same canalized output,
\begin{equation}
  \eta = a^2 + (1-a)^2,
\end{equation}
and the probability $\eta_0$ that the last input has the same canalized output
as the default output,
\begin{equation}
  \eta_0 = ad + (1-a)(1-d).
\end{equation}
Using this, we can write the probability $\lambda_i$ that the output will be
flipped if input $i$ is flipped,
\begin{equation}\label{eq:li}
  \begin{split}
  \lambda_i &=\, (1-\gamma(b^*))^i\left[(1-(1-\gamma(b^*))^{k-i-1})(1-\eta)\, + \right.\\
            &\relphantom{=} \phantom{(1-\gamma(b^*))^i\bigl[} \left. (1-\gamma(b^*))^{k-i-1}(1-\eta^0)\right],
  \end{split}
\end{equation}
which accounts for the probability that all the inputs $j<i$ are not canalized,
and that the output of $i$ is different than any input $j>i$ which may be
canalized. The sensitivity value $\lambda = \sum_i \lambda_i$ amounts then
simply to,
\begin{equation}\label{eq:lambda}
  \lambda = (1-\eta) \frac{1-(1-\gamma(b^*))^k}{\gamma(b^*)} +
  k (1 - \gamma(b^*))^{k-1}(\eta - \eta^0).
\end{equation}
The situation where $s_d = 1 - s_{k-1}$ can be obtained by setting $d=1-a$, as
before, and making the substitution $\eta^0 \to \eta^0(1-\delta_{i,k-1})$ in
Eq.~\ref{eq:li}, which yields
\begin{equation}\label{eq:lambdas}
  \begin{split}
  \lambda^s &= (1-\eta) \frac{1-(1-\gamma(b^*))^k}{\gamma(b^*)} + \\
            &\relphantom{=} (1 - \gamma(b^*))^{k-1}\left[k(\eta - 1) + (1-\eta^0)(k-1) + 1\right].
\end{split}
\end{equation}

\begin{figure}[hbt!]
  \includegraphics*[width=\columnwidth]{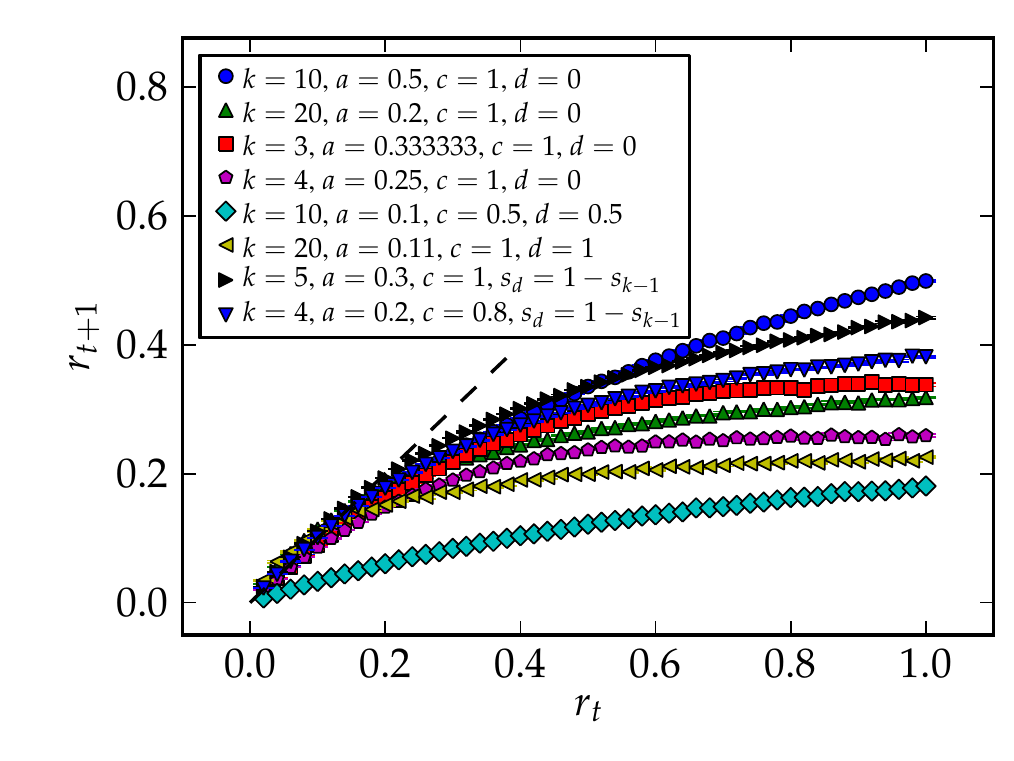}\\
  \includegraphics*[width=\columnwidth]{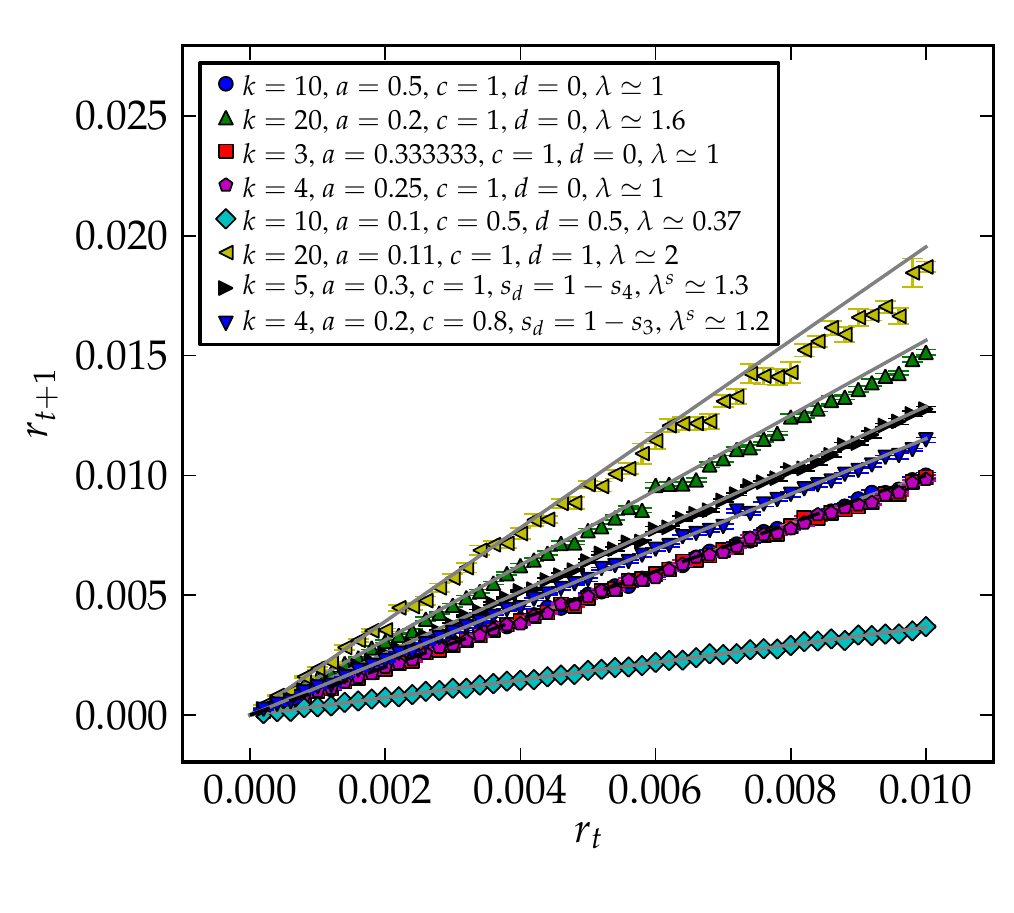}
  \caption{\label{fig:derrida} Fraction of perturbed nodes $r_{t+1}$ after one
    time step, as a function of an initial fraction of random perturbed nodes
    $r_t$, for parameter values marked as in Figs.\ref{fig:lambda} and
    \ref{fig:lambdas}, and shown in the legend. The solid lines are
    slopes of the form $r_{t+1} = \lambda r_t$, with $\lambda$ or $\lambda^s$
    calculated according to Eqs.\ref{eq:lambda} or \ref{eq:lambdas},
    respectively.}
\end{figure}

We note that $\lambda^s \ge \lambda$ for small values of $k$, but for $k\gg1$ we
have $\lambda^s \sim \lambda$, and this value approaches $\lim_{k\to\infty}
\lambda = (1 - \eta)/\gamma(b^*)$, for $\gamma(b^*) > 0$. This means that in the
limit of large $k$, NCFs tend to be much more stable than randomly chosen
functions, which have $\lambda=k/2$.

In Figs.~\ref{fig:lambda} and~\ref{fig:lambdas} are the phase diagrams for
several parameter values. We note that these diagrams are symmetric in respect
to the appropriate parameter transformations, such as $(c,d) \leftrightarrow
(1-c, 1-d)$.  For $d=0$ and $c=1$, we observe that the critical line is composed
of the $a=1/2$ line plus the $a=1/k$ line, which corresponds to the critical
values of $a$ where $b^* > 0$, as discussed previously. We note that, since
$\lambda \cong (1 - \eta)/\gamma(b^*)$ for $k\gg 1$, there is a universal
critical line at $a=1/2$ for which $\eta = 1/2$ and $\gamma(b^*) \cong 1/2$, and
thus $\lambda \cong 1$, independent of $c$ and $d$. This is interesting, since
it means that the most \emph{entropic} situation $a=c=d=1/2$ leads to a critical
network, if $k$ is not too small, and either changing $c$ or $d$ leaves the
value of $\lambda$ unmodified. Thus criticality can be attained for a large
number of possible configurations, which is maybe a reason why this class of
functions are favoured biologically. However, there are large portion of the
configuration space where $\lambda > 1$, and the dynamics finds itself in the
chaotic phase. Additionally, there are significant regions where the fixed point
of $b^*$ is unstable ans is replaced by a period-2 oscillation, as discussed
previously. In this situation (which are marked in white in
Figs.\ref{fig:lambda} and \ref{fig:lambdas}), the value of $\lambda$ is not
meaningful, since it supposes that $b^*$ is a stable fixed point, and the
networks are neither in the frozen nor in the chaotic phase.

The values of $\lambda$ and $\lambda^s$ in Eqs.~\ref{eq:lambda}
and~\ref{eq:lambdas} can be verified empirically, by constructing RBNs and
obtaining the so-called Derrida plots~\cite{derrida_evolution_1986}, as shown in
Fig.~\ref{fig:derrida}. These plots show the normalized hamming distance
$r_{t+1}$ at time $t+1$ between two identical copies of the network, after only
one of them had a random fraction of $r_t$ nodes flipped at time $t$. As can be
seen on the top of Fig.~\ref{fig:derrida}, the different networks (even those
with the same value of $\lambda$) have a different perturbation
propagation. However, if we only consider small values of $r_t$, these curves
are matched quite exactly by slopes of type $r_{t+1} = \lambda r_t$, as can be
seen on the bottom of Fig.~\ref{fig:derrida}.

\section{Conclusion}\label{sec:conclusion}

We have obtained the phase diagram of random Boolean networks with nested
canalizing functions. Using the annealed approximation, we have analytically
calculated the fraction $b_t$ of nodes with value one, and the sensitivity
$\lambda$ of the network to small perturbations. We compared the results with
numerical realizations of quenched networks, which have shown an excellent
agreement. We have seen that the steady state value $b^* = \lim_{t\to\infty}b_t$
displays two interesting features for some parameter combinations: 1. For a
probability $c=1$ that the canalizing value of a random input is one, and a
probability $d=0$ that the default output of a random input is one (and
symmetrically for $d=1$ and $c=0$), there is a second-order transition from
$b^*=0$ to $b^*>0$ at critical fraction $a=1/k$ of inputs which are activators;
and 2. If $|d-c|$ is small enough, $b_t$ will oscillate between two values of
$b^*$, up to a value of $a$ for which the fixed point of $b_t$ will become
stable again. A similar oscillation is also observed in RBNs with threshold
functions~\cite{greil_kauffman_2007}.

We have also observed that the phase diagram has large regions where
$\lambda>1$, and thus the system is in the chaotic phase. This contradicts the
claim by Kauffman~\textit{et al.}~\cite{kauffman_genetic_2004} that this class
of functions always leads to $\lambda<1$. The results
in~\cite{kauffman_genetic_2004} were obtained for specific in-degree
distributions, and a selection of the canalizing values according to specific
probabilities. These probabilities were chosen so that they match the
experimental data they analysed, but for which no other explanation was
offered. This distribution was then extrapolated to obtain the entire phase
diagram, which resulted only in values of $\lambda<1$. In this work, the
functions were chosen with different probabilities, according to the simple
parameters $a$, $c$ and $d$, which resulted in a phase diagram with large
portions where $\lambda>1$. Therefore, since there seems to be no reason to
believe the specific parametrization in~\cite{kauffman_genetic_2004} has a
general character, it should not be concluded that the existence of NCFs always
leads to $\lambda<1$. On the other hand, the values of $\lambda$ for RBNs with
NCFs are much \emph{smaller} than RBNs with all functions chosen with equal
probability. Thus the conclusion in~\cite{kauffman_genetic_2004} that NCFs
convey more stability to the system seem well justified, and it may indeed be a
reason why these functions are observed in real biological systems.

This work has been supported by the DFG under Contract No. Dr300/5-1.

\bibliographystyle{epj}
\bibliography{bib}

\end{document}